\documentclass[review,12pt]{elsarticle}

\usepackage{amssymb}
\usepackage{amsthm}
\usepackage{framed} 
\usepackage{amsmath,color}
\usepackage{mathrsfs}
\usepackage{graphicx}
\usepackage{epstopdf}
\usepackage{float}
\usepackage{caption}
\usepackage{subcaption}
\usepackage{bm}
\usepackage{bbm}
\usepackage{mathrsfs}
\usepackage{cleveref}
\usepackage{soul}
\usepackage{upgreek} 
\usepackage{accents}
\usepackage{color,soul} 
\usepackage{color} 
\usepackage{nomencl} 
\biboptions{sort&compress}
\soulregister\citep7 
\soulregister\citet7 
\soulregister\citealp7 
\newsavebox{\measurebox} 
\usepackage{titlesec} 
\usepackage{tabu} 
\usepackage{longtable}
\usepackage{framed} 
\usepackage{nomencl} 
\makenomenclature 

\journal{Engineering Fracture Mechanics}

\makeatletter
\def\@author#1{\g@addto@macro\elsauthors{\normalsize%
    \def\baselinestretch{1}%
    \upshape\authorsep#1\unskip\textsuperscript{%
      \ifx\@fnmark\@empty\else\unskip\sep\@fnmark\let\sep=,\fi
      \ifx\@corref\@empty\else\unskip\sep\@corref\let\sep=,\fi
      }%
    \def\authorsep{\unskip,\space}%
    \global\let\@fnmark\@empty
    \global\let\@corref\@empty  
    \global\let\sep\@empty}%
    \@eadauthor={#1}
}
\makeatother

\titleformat{\paragraph}
{\normalfont\normalsize\itshape}{\theparagraph}{1em}{}
\titlespacing*{\paragraph}
{0pt}{3.25ex plus 1ex minus .2ex}{1.5ex plus .2ex}

\begin{document}

\begin{frontmatter}



\title{A mechanism-based gradient damage model for metallic fracture}


\author{Siamak S. Shishvan \fnref{Tabriz}}

\author{Saeid Assadpour-asl \fnref{Tabriz}}

\author{Emilio Mart\'{\i}nez-Pa\~neda\corref{cor1}\fnref{IC}}
\ead{e.martinez-paneda@imperial.ac.uk}

\address[Tabriz]{Department of Structural Engineering, University of Tabriz, P.O. Box 51666-16471, Tabriz, Iran}

\address[IC]{Department of Civil and Environmental Engineering, Imperial College London, London SW7 2AZ, UK}

\cortext[cor1]{Corresponding author.}

\begin{abstract}
A new gradient-based formulation for predicting fracture in elastic-plastic solids is presented. Damage is captured by means of a phase field model that considers both the elastic and plastic works as driving forces for fracture. Material deformation is characterised by a mechanism-based strain gradient constitutive model. This non-local plastic-damage formulation is numerically implemented and used to simulate fracture in several paradigmatic boundary value problems. The case studies aim at shedding light into the role of the plastic and fracture length scales. It is found that the role of plastic strain gradients is two-fold. When dealing with sharp defects like cracks, plastic strain gradients elevate local stresses and facilitate fracture. However, in the presence of non-sharp defects failure is driven by the localisation of plastic flow, which is delayed due to the additional work hardening introduced by plastic strain gradients.
\end{abstract}


\end{frontmatter}


\begin{framed}
\nomenclature{$\Omega$}{domain of the solid}
\nomenclature{$\mathbf{n}$}{surface unit normal}
\nomenclature{$\mathbf{u}$}{displacement field vector}
\nomenclature{$\bm{\varepsilon}, \, \bm{\varepsilon}^e, \bm{\varepsilon}^p$}{total, elastic and plastic strain tensors}
\nomenclature{$\phi$}{Phase field order parameter}
\nomenclature{$\ell_f$}{Phase field fracture length scale}
\nomenclature{$G_c$}{material toughness}
\nomenclature{$\Gamma$}{discontinuous surface}
\nomenclature{$\gamma$}{crack density functional}
\nomenclature{$\bm{\sigma}$}{Cauchy stress tensor}
\nomenclature{$\mathbf{T}$}{traction vector}
\nomenclature{$\omega$}{scalar micro-stress}
\nomenclature{$\bm{\upxi}$}{micro-stress vector}
\nomenclature{$\psi, \, \psi^e, \, \psi^p$}{total, elastic and plastic strain energy densities}
\nomenclature{$\varphi$}{fracture energy density}
\nomenclature{$\lambda$}{first Lame parameter}
\nomenclature{$\kappa$}{ill-conditioning parameter}
\nomenclature{$c_w$}{phase field scaling constant}
\nomenclature{$a$}{crack length}
\nomenclature{$b$}{Burgers vector}
\nomenclature{$\bm{B}_i^u$}{nodal strain-displacement matrices}
\nomenclature{$\mathbf{C}_{ep}$}{elastic-plastic material Jacobian}
\nomenclature{$E$}{Young's modulus}
\nomenclature{$K_I$}{Mode I stress intensity factor}
\nomenclature{$K_0$}{Reference stress intensity factor}
\nomenclature{$\ell_p$}{Strain gradient plasticity length scale}
\nomenclature{$M$}{Taylor's factor}
\nomenclature{$m$}{rate sensitivity exponent}
\nomenclature{$N_i$}{nodal shape functions}
\nomenclature{$\bm{N}_i$}{shape functions nodal interpolation matrices}
\nomenclature{$N$}{strain hardening exponent}
\nomenclature{$R_0$}{reference fracture process zone length}
\nomenclature{$\bar{r}$}{Nye's factor}
\nomenclature{$r$,\, $\theta$}{polar coordinates}
\nomenclature{$u , \, v$}{horizontal and vertical components of the displacement field}
\nomenclature{$\varepsilon^p$}{equivalent plastic strain}
\nomenclature{$\eta^p$}{effective plastic strain gradient}
\nomenclature{$\mu$}{shear modulus}
\nomenclature{$\nu$}{Poisson's ratio}
\nomenclature{$\rho$}{total dislocation density}
\nomenclature{$\rho_G$}{density of Geometrically Necessary Dislocations (GNDs)}
\nomenclature{$\rho_S$}{density of Statistically Stored Dislocations (SSDs)}
\nomenclature{$\bm{\eta}^p$}{plastic strain gradient tensor}
\nomenclature{$\sigma_Y$}{initial yield stress}
\nomenclature{$\sigma_e$}{effective stress}
\nomenclature{$\hat{\sigma}$}{material strength}
\nomenclature{$\mathcal{H}$}{phase field history field}
\nomenclature{$\sigma_{flow}$}{tensile flow stress}
\nomenclature{$\tau$}{shear flow stress}
\printnomenclature
\end{framed}


\section{Introduction}
\label{Sec:Introduction}

It has been 100 years since Griffith's seminal work \cite{Griffith1920} started a century of fracture mechanics research and, undoubtedly, some areas within this broad discipline have achieved a high degree of maturity. As a consequence, fracture mechanics is nowadays an essential tool for ensuring durability, efficiency and safety across a wide range of sectors and applications. However, many challenges remain and, despite the progress achieved, the discipline continues to attract a notable degree of interest from academics and practitioners \cite{Anderson2005,Kendall2021}. Well-known longstanding issues are related to multi-physics problems and applications involving subcritical crack growth. With growing interest for fail-safe and damage-tolerance approaches to design comes the need to develop robust computational methods capable of predicting the nucleation and growth of defects. In this regard, the phase field fracture method has attracted particular attention over the last decade. Phase field methods aim at substituting the boundary conditions at an interface by a partial differential equation for the evolution of an auxiliary (phase) field. Thus, the problem is solved by integrating a set of partial differential equations for the whole system, avoiding the explicit treatment of the interface conditions. Phase field methods are finding ever increasing applications, from microstructural evolution \cite{Chen2002} to corrosion damage \cite{JMPS2021}. In the case of fracture problems, the phase field order parameter $\phi$ implicitly describes the crack-solid interface. The phase field can be thought of as a damage variable, taking a value of $\phi=0$ at intact material points and of $\phi=1$ when the material point is fully cracked, with a smooth variation in-between. Earlier efforts were focused on ideally elastic solids and consequently the phase field evolution equation was grounded on Griffith's energy balance \cite{Bourdin2008,Kuhn2010,Miehe2010a,Linse2017,PTRSA2021}. The method proved to be a success in modelling brittle fracture and the number of applications soared: composite materials \cite{Quintanas-Corominas2019,CST2021,Bui2021}, functionally graded materials \cite{CPB2019,Kumar2021,DT2020}, shape memory alloys \cite{CMAME2021}, hyperelastic solids \cite{Mandal2020a,Peng2020}, hydraulic fracture \cite{Xia2017,Chen2020a}, fatigue damage \cite{Alessi2018c} and hydrogen-embrittled alloys \cite{CMAME2018,TAFM2020c}, just to name some - see Refs. \cite{Wu2020} for a review.\\

In recent years there has been an increasing interest in extending the success of phase field methods to the modelling of fracture in elastic-plastic solids. Models have been proposed for both brittle fracture under small scale yielding conditions \cite{Duda2015,JMPS2020} and ductile damage \cite{Ambati2015a,Borden2016,Miehe2016b,Miehe2016c,CS2020}; see Ref. \cite{Alessi2018} for a critical overview. The majority of these models base the constitutive behaviour of the solid on von Mises plasticity theory. However, conventional continuum models, such as von Mises plasticity, fail to capture the dislocation hardening mechanisms governing crack tip mechanics \cite{Wei1997,IJP2016}. Namely, Geometrically Necessary Dislocations (GNDs) arise due to the need to accommodate the large plastic strain gradients that develop in the vicinity of the crack tip. This extra storage of dislocations elevates local strength due to mechanisms such as forest hardening or due to long range back-stresses associated with the stored elastic energy of GNDs. Strain gradient plasticity models have been developed to account for the role of plastic strain gradients and the associated dislocation hardening mechanisms at the continuum level \cite{Fleck1994,Gao1999,Gurtin2005}. The analysis of stationary crack tip fields using strain gradient plasticity models reveals much higher stresses than those predicted with conventional plasticity, with this stress elevation being sustained over tens of microns ahead of the crack \cite{Komaragiri2008,IJSS2015}. Thus, it is necessary to incorporate strain gradient plasticity formulations into fracture modelling to appropriately characterise the small scales associated with crack tip deformation. Non-local and gradient approaches to damage in elastic-plastic solids have been presented without incorporating the phase field formalism. Examples include integral-type approaches \cite{Bazant2003}, implicit and explicit gradient models \cite{Engelen2002,Peerlings2002}, micromorphic theories \cite{Brepols2017,Ling2018,Forest2019} and energy-based approaches \cite{Lancioni2015}. Gradient-damage models have been compared with phase field fracture approaches in the context of elastic solids, showing that the structure of the phase field balance equation prevents the nonphysical damage zone broadening observed in gradient-damage models \cite{DeBorst2016a,Mandal2019a}. Such a comparison has not been presented yet in the context of elastic-plastic fracture.\\

In this work, we present a new phase field fracture model for elastic-plastic solids that incorporates the role of dislocation hardening and plastic strain gradients through a mechanistic formulation based on Taylor's \cite{Taylor1938} dislocation model. The potential and predictions of the model are showcased by addressing a number of boundary value problems of particular interest. Emphasis is placed on investigating the role of the plastic and phase field length scales, which result from the non-locality of the model in regard to plastic strain and damage gradients. The remainder of this manuscript is organised as follows. The new dislocation-based elastic-plastic phase field fracture theory presented is described in Section \ref{Sec:Theory}. Details of the finite element implementation are given in Section \ref{Sec:NumModel}. Representative results are shown in Section \ref{Sec:Results}. Spanning different triaxiality conditions, the predictions of the model are examined by modelling crack propagation in a boundary layer model, a compact tension specimen and an asymmetric double-notched specimen. The manuscript ends with concluding remarks in Section \ref{Sec:Concluding remarks}. 

\section{Theory}
\label{Sec:Theory}

The implicitly multi-scale damage model for elastic-plastic fracture presented stands on a Taylor-based strain gradient plasticity formulation and a phase field formulation for elastic-plastic fracture. We shall first present the kinematics (Section \ref{Sec:Kinematics}) and balance equations (Section \ref{Sec:PVW}) of the coupled problem, and proceed to make specific constitutive choices (Section \ref{Sec:ConstitutiveTheory}). The theoretical framework presented herein refers to an elastic-plastic solid occupying an arbitrary domain $\Omega \subset {\rm I\!R}^n$ $(n \in[1,2,3])$, with an external boundary $\partial \Omega\subset {\rm I\!R}^{n-1}$, on which the outwards unit normal is denoted as $\mathbf{n}$.

\subsection{Kinematics}
\label{Sec:Kinematics}

As elaborated below, we employ a first-order approach in the modelling of gradient effects. Moreover, we restrict our attention to small strains and isothermal conditions. As a consequence, the primal kinematic variables of the problem are the displacement field vector $\mathbf{u}$ and the damage phase field $\phi$. The strain tensor $\bm{\varepsilon}$ is defined as
\begin{equation}
    \bm{\varepsilon} = \frac{1}{2}\left(\nabla\mathbf{u}^T+\nabla\mathbf{u}\right) \, ,
\end{equation}

\noindent and it can be additively decomposed into its elastic $\bm{\varepsilon}^e$ and plastic $\bm{\varepsilon}^p$ parts as
\begin{equation}
    \bm{\varepsilon}=\bm{\varepsilon}^e + \bm{\varepsilon}^p \, .
\end{equation}

The nucleation and subsequent propagation of cracks are described by using a smooth continuous scalar \emph{phase field} $\phi \in [0;1]$. The phase field describes the degree of damage in each material point, as in continuum damage mechanics approaches. Here, we assume that $\phi=0$ corresponds to the case where the material point is in its intact state, while $\phi=1$ denotes the material that is fully broken. Since $\phi$ is smooth and continuous, discrete cracks are represented in a diffuse fashion. The smearing of cracks is controlled by a phase field length scale $\ell_f$, which appears due to dimensional consistency and makes the model non-local, guaranteeing mesh objectivity. The purpose of the diffuse phase field representation is to introduce the following approximation of the fracture energy over a discontinuous surface $\Gamma$: 
\begin{equation}
    \Phi=\int_{\Gamma} G_c \, \text{d}S \approx \int_\Omega G_c\gamma(\phi,\nabla\phi) \, \text{d}V, \hspace{1cm} \text{for } \ell_f\rightarrow 0,
\end{equation}

\noindent where $\gamma$ is the so-called crack surface density functional and $G_c$ is the material fracture energy \cite{Griffith1920,Irwin1956}. The latter provides a measure of the toughness of the solid, as first presented by Griffith \cite{Griffith1920} and Irwin \cite{Irwin1956} for elastic solids and later extended to account for inelastic energy dissipation by Orowan \cite{Orowan1948}.

\subsection{Principle of virtual work and balance of forces}
\label{Sec:PVW}

We proceed to derive the balance equations using the principle of virtual work. The Cauchy stress tensor $\bm{\sigma}$ is introduced, which is work conjugate to the strain tensor $\bm{\varepsilon}$. Correspondingly, a traction $\mathbf{T}$ is defined on the boundary $\partial\Omega$, which is work conjugate to the displacements $\mathbf{u}$. Regarding damage, we introduce a scalar stress-like quantity $\omega$, which is work conjugate to the phase field $\phi$, and a phase field micro-stress vector $\bm{\upxi}$ that is work conjugate to the gradient of the phase field $\nabla\phi$. The phase field is assumed to be driven solely by the energy being released by the solid; i.e., no external traction is associated with $\phi$. Accordingly, in the absence of body forces, the principle of virtual work is given by:
\begin{equation}\label{eq:PVW}
 \int_\Omega \big\{ \bm{\sigma}:\delta\bm{\varepsilon}  + \omega\delta\phi+\bm{\upxi} \cdot \delta \nabla \phi
    \big\} \, \text{d}V =  \int_{\partial \Omega} \left( \mathbf{T} \cdot \delta \mathbf{u} \right) \, \text{d}S \, ,
\end{equation}
\noindent where $\delta$ denotes a virtual quantity. This equation must hold for an arbitrary domain $\Omega$ and for any kinematically admissible variations of the virtual fields. Hence, by application of the Gauss divergence theorem, the local force balances are given by: 
\begin{equation}
    \begin{split}
        &\nabla\cdot\bm{\sigma}=0  \\
        &\nabla\cdot\bm{\upxi}-\omega =0
    \end{split}\hspace{2cm} \text{in } \,\, \Omega,\label{eq:balance}
\end{equation}

\noindent with natural boundary conditions: 
\begin{equation}
    \begin{split}
        \bm{\sigma}\cdot\mathbf{n}=\mathbf{T} \\
         \bm{\upxi} \cdot \mathbf{n}=0 
    \end{split} \hspace{2cm} \text{on } \,\, \partial\Omega.\label{eq:balance_BC}
\end{equation}

\subsection{Constitutive relations}
\label{Sec:ConstitutiveTheory}

Now, we shall make constitutive choices for the deformation and fracture problems. The Taylor-based strain gradient plasticity theory adopted for the constitutive deformation behaviour of the solid is presented first (Section \ref{Sec:MSGplasticity}). Then, the phase field choices of crack density functional, fracture driving force and degradation function are described in Section \ref{Sec:PhaseField}. 

\subsubsection{Mechanism-based Strain Gradient Plasticity}
\label{Sec:MSGplasticity}

The elastic-plastic response of the solid is described by the so-called mechanism-based strain gradient (MSG) plasticity theory \citep{Gao1999,Qiu2003}. The aim is to capture the role of GNDs in the mechanics of crack initiation and growth. MSG plasticity theory is grounded on Taylor's dislocation model \cite{Taylor1938} where the shear flow stress $\tau$ is defined in terms of the dislocation density $\rho$, the shear modulus $\mu$ and the Burgers vector $b$ as
\begin{equation}\label{Eq1MSG}
\tau = \alpha \mu b \sqrt{\rho} \, .
\end{equation}

\noindent Here, $\alpha$ is an empirical coefficient that is assumed to be equal to 0.5. The dislocation density $\rho$ is additively decomposed into the density of statistically stored dislocations (SSDs), $\rho_S$, and the density of GNDs, $\rho_G$, as
\begin{equation}\label{Eq2MSG}
\rho = \rho_S + \rho_G \, .
\end{equation}

Defining $\overline{r}$ as Nye's factor, which is assumed to be equal to 1.9 for fcc polycrystals, the GND density $\rho_G$ is related to the effective plastic strain gradient $\eta^{p}$ by
\begin{equation}\label{Eq3MSG}
\rho_G = \overline{r}\frac{\eta^{p}}{b} \, .
\end{equation}

\noindent The effective plastic strain gradient $\eta^{p}$ is defined by considering three invariants of the plastic strain gradient tensor, as follows
\begin{equation}
\eta^{p}=\sqrt{c_1 \bm{\eta}^{p} \bm{\eta}^{p} + c_2 \bm{\eta}^{p} \bm{\eta}^{p} + c_3 \bm{\eta}^{p} \bm{\eta}^{p}} \, .
\end{equation}

The coefficients have been determined to be $c_1=0$, $c_2=1/4$ and $c_3=0$ from three dislocation models for bending, torsion and void growth \cite{Gao1999} leading to
\begin{equation}
\eta^{p}=\sqrt{\frac{1}{4}\bm{\eta}^{p} \bm{\eta}^{p}}
\end{equation}

\noindent where the components of the strain gradient tensor are given by
\begin{equation}
\eta^{p}_{ijk}= \varepsilon^{p}_{ik,j}+\varepsilon^{p}_{jk,i}-\varepsilon^{p}_{ij,k}
\end{equation}

In Taylor's dislocation model, the tensile flow stress $\sigma_{flow}$ is the product of the shear flow stress $\tau$ and Taylor's factor $M$, which is equal to 3.06 for fcc metals, so as
\begin{equation}\label{Eq4MSG}
\sigma_{flow} =M\tau
\end{equation}

\noindent Rearranging (\ref{Eq1MSG}-\ref{Eq3MSG}) and substituting into (\ref{Eq4MSG}) renders
\begin{equation}\label{Eq5MSG}
\sigma_{flow} =M\alpha \mu b \sqrt{\rho_{S}+\overline{r}\frac{\eta^{p}}{b}}
\end{equation}

The SSD density $\rho_{S}$ can be readily determined from (\ref{Eq5MSG}) knowing the relation in uniaxial tension ($\eta^p=0$) between the flow stress and the material stress-strain curve,
\begin{equation}\label{Eq6MSG}
\rho_{S} = \left[ \frac{\sigma_{ref}f(\varepsilon^{p})}{M\alpha \mu b} \right]^2
\end{equation}

\noindent Here, $\sigma_{ref}$ is a reference stress and $f$ is a non-dimensional function of the plastic strain $\varepsilon^{p}$, as given by the uniaxial stress-strain curve. Substituting into (\ref{Eq5MSG}), the flow stress $\sigma_{flow}$ reads
\begin{equation}\label{EqSflow}
\sigma_{flow} =\sigma_{ref} \sqrt{f^2(\varepsilon^{p})+\ell_p \eta^{p}}
\end{equation}

\noindent where $\ell_p$ is the intrinsic material plastic length parameter, which enters the constitutive equation as a result of dimensional consistency. The value of $\ell_p$ is commonly obtained by fitting micro-scale experiments \cite{IJES2020}. The model recovers the conventional von Mises plasticity solution when $\ell_p=0$. It remains to make constitutive choices for $\sigma_{ref}$ and $f(\varepsilon^{p})$; unless otherwise stated, we here assume the following power-law hardening behaviour:
\begin{equation}\label{eq:HardeningPowerLaw}
\sigma = \sigma_{ref} f \left( \varepsilon^{p} \right) = \sigma_Y \left( \frac{E}{\sigma_Y} \right)^N \left( \varepsilon^p + \frac{\sigma_Y}{E} \right)^N = \sigma_Y \left( 1 + \frac{E \varepsilon^p}{\sigma_Y} \right)^N
\end{equation}

\noindent where $\sigma_Y$ is the initial yield stress and $N$ is the strain hardening exponent. This choice of hardening law leads to a power law relation between $\rho_S$ and $\varepsilon^p$; this description can be enriched to account for the processes of multiplication and annihilation of SSDs (see, e.g., \cite{Ma2006a,Petryk2016}).

\subsubsection{Phase field fracture}
\label{Sec:PhaseField}

Variational phase field fracture models differ on their choices for the crack density functional, the fracture driving force and the degradation function. Here, we outline our choices, which generally correspond to those of the standard or \texttt{AT2} phase field model \cite{Bourdin2000} and include a driving force for fracture that considers both elastic and plastic strain energy densities. We start by defining the total potential energy of the solid as,
\begin{equation}\label{eq:TotalPotentialEnergy0}
W \left( \bm{\varepsilon} \left( \mathbf{u} \right), \, \phi, \,  \nabla \phi \right) = \psi \left( \bm{\varepsilon} \left( \mathbf{u} \right), \, g \left( \phi \right) \right)  +  \varphi \left( \phi, \, \nabla \phi \right)
\end{equation}

\noindent where $\psi$ is the strain energy density and $\varphi$ is the fracture energy density. The total strain energy density $\psi$ can be additively decomposed into its elastic $\psi^e$ and plastic $\psi^p$ parts, and thus computed as follows:
\begin{equation}\label{eq:StrainEnergyDensity}
    \psi = \psi^e \left(\bm{\varepsilon}^e \right) + \psi^p \left( \bm{\varepsilon}^p \right)= \frac{1}{2} \lambda \left[ \text{tr} \left( \bm{\varepsilon}^e \right) \right]^2 + \mu \, \text{tr} \left[  \left( \bm{\varepsilon}^e \right)^2 \right] + \int_0^t \left( \bm{\sigma} : \dot{\bm{\varepsilon}}^p \right) \text{d} t \, .
\end{equation}

\noindent The Cauchy stress tensor is then defined as $\bm{\sigma}=\partial_{\bm{\varepsilon}} \psi$. In this work, the fracture driving force is taken to be equal to the total strain energy density, following Miehe \textit{et al.} \cite{Miehe2016b} and Borden \textit{et al.} \cite{Borden2016}. Other approaches have also been considered in the literature. For example, only the elastic, stored energy is assumed to be available for crack growth in Refs. \cite{Duda2015,JMPS2020}. A suitable choice is not straightforward. In quasi-static experiments at room temperature, most of the plastic work is dissipated into heat \cite{Taylor1934} and is not available to be converted into fracture energy. However, an energy balance \emph{\`{a} la} Griffith is not suitable for fracture processes involving significant plasticity \cite{Hutchinson1983}. For example, the assumption of a continuous temperature at the crack tip is no longer valid as local plastic flow constitutes a source of heat \cite{Gurtin1979}. Thus, in the case of fracture in the presence of significant plastic flow, the phase field balance law weakens its connection with Griffith's (and Orowan's) thermodynamics and becomes more phenomenological in nature. Unlike some classes of plasticity damage models, the contribution of the elastic work to fracture is not neglected, and this allows using the same framework for predicting ductile damage and quasi-cleavage fracture, as observed in (e.g.) embrittled alloys.\\

The strain energy density of the solid diminishes with increasing damage through the degradation function $g \left( \phi \right)$, which must fulfill the following conditions:
\begin{equation}
g \left( 0 \right) =1 , \,\,\,\,\,\,\,\,\,\, g \left( 1 \right) =0 , \,\,\,\,\,\,\,\,\,\, g' \left( \phi \right) \leq 0 \,\,\, \text{for} \,\,\, 0 \leq \phi \leq 1 \, .
\end{equation}

\noindent Here, we choose to adopt the widely used quadratic degradation function such that
\begin{equation}
    g \left( \phi \right) = \left( 1 - \phi \right)^2 
\end{equation}

We proceed to formulate the fracture energy density as,
\begin{align}
    \varphi \left( \phi, \, \nabla \phi \right) = G_c \gamma(\phi, \nabla\phi) = G_c \dfrac{1}{4c_w\ell_f}\left( w(\phi) + \ell_f^2 |\nabla\phi|^2\right) \, .
\end{align}

\noindent where $\ell_f$ is the phase field length scale, $c_w$ is a scaling constant and $w(\phi)$ is the geometric crack function. The constitutive choice for the geometric crack function must satisfy the following conditions:
\begin{equation}
w \left( 0 \right) =0 , \,\,\,\,\,\,\,\,\,\, w \left( 1 \right) =1 , \,\,\,\,\,\,\,\,\,\, w' \left( \phi \right) \geq 0 \,\,\, \text{for} \,\,\, 0 \leq \phi \leq 1 \, .
\end{equation}

\noindent The scaling constant $c_w$ can be derived from the geometric crack function:
\begin{equation}
    c_w = \int_0^1\sqrt{w(\zeta)} \, \text{d}\zeta \, .
\end{equation}

\noindent Here, we assume $w(\phi)=\phi^2$ and $c_w=1/2$, which are constitutive choices associated with the so-called \texttt{AT2} phase field model. Without loss of generality, one can re-formulate the total potential energy of the solid (\ref{eq:TotalPotentialEnergy0}) as,
\begin{equation}\label{eq:Free_energy}
     W =  {g(\phi)} \psi  + \frac{G_c}{4 c_w }  \left(\frac{1}{ \ell_f} {w(\phi)}+\ell_f |\nabla \phi|^2\right) \, 
\end{equation}

\noindent from which the fracture micro-stress variables $\omega$ and $\bm{\upxi}$ can be readily derived as follows. While the scalar micro-stress $\omega$ reads
\begin{equation}\label{eq:consOmega}
    \omega = \dfrac{\partial W}{\partial\phi} = {g^{\prime}(\phi)} \psi +\frac{G_c}{4c_w \ell_f}  w^{\prime}(\phi) \, ,
\end{equation}

\noindent the phase field micro-stress vector $\bm{\upxi}$ is given by,
\begin{equation}\label{eq:consXi}
    \bm{\upxi} = \dfrac{\partial W}{\partial\nabla\phi} = \frac{\ell_f}{2c_w} G_{\mathrm{c}} \nabla \phi \, .
\end{equation}

\noindent Inserting Eqs. (\ref{eq:consOmega})-(\ref{eq:consXi}) into the phase field balance equation (\ref{eq:balance}b), the phase field evolution law can be reformulated as:
\begin{equation}\label{eq:PhaseFieldStrongForm}
  \frac{G_c}{2c_w}  \left( \frac{w^{\prime}(\phi)}{2 \ell_f} - \ell_f \nabla^2 \phi \right) + {g^{\prime}(\phi)} \psi = 0  
\end{equation}

Finally, we note that the phase field length scale $\ell_f$ can be related to the material strength \cite{Borden2012,Tanne2018}. Consider a simple 1D problem, such as the tensile testing of a smooth linear elastic bar; in the phase field evolution law $\nabla \phi=0$ and $\psi=E \varepsilon^2/2$ and solving for $\phi$ in (\ref{eq:PhaseFieldStrongForm}) renders:
\begin{equation}
     \phi = \frac{E \varepsilon^2 \ell_f}{G_c  + E \varepsilon^2 \ell_f} \, .
\end{equation}

\noindent Hence with $\sigma_0$ denoting the undamaged stress, the effective stress $\sigma=\left( 1 - \phi \right)^2 \sigma_0$ reaches a maximum at
\begin{equation}\label{eq:effectiveS}
    \hat{\sigma} = \left( \frac{27E G_c }{256 \ell_f} \right)^{1/2} \, .
\end{equation}

\section{Numerical implementation}
\label{Sec:NumModel}

We proceed to describe the numerical implementation of the coupled deformation-fracture theory presented in Section \ref{Sec:Theory}. First, in Section \ref{eq:MSG_FEM}, the details of the implementation of the mechanism-based strain gradient (MSG) plasticity constitutive material model are presented. Numerical aspects related to the phase field problem are described in Section \ref{Sec:IrreversibilitySplit}; namely, the need to enforce irreversibility and to prevent damage under compression. Finally, in Section \ref{Sec:FEdiscretization}, the weak form is discretised and the stiffness matrices and residuals of the coupled problem are derived. 

\subsection{First order MSG plasticity implementation}
\label{eq:MSG_FEM}

We choose to implement the MSG plasticity constitutive model using a first-order scheme. First-order and second-order implementations of MSG plasticity predict identical results over their physical domain of validity \cite{Huang2004a,TAFM2017} and the use of a lower order approach circumvents convergence issues associated with the use of higher order terms \cite{Hwang2003}. A viscoplastic approach shall be adopted to achieve a first-order, self-consistent implementation of MSG plasticity. Recall that the Taylor dislocation model defines the flow stress $\sigma_{flow}$ to be dependent on both the equivalent plastic strain $\varepsilon^p$ and the effective plastic strain gradient $\eta^p$, see (\ref{EqSflow}). It then follows that,
\begin{equation}
\dot{\sigma}_{flow}=\frac{\partial \sigma_{flow}}{\partial \varepsilon^p} \dot{\varepsilon}^p + \frac{\partial \sigma_{flow}}{\partial \eta^p} \dot{\eta}^p \, .
\end{equation}

\noindent Thus, as noted by Huang \textit{et al.} \cite{Huang2004a}, for a plastic strain rate $\dot{\bm{\varepsilon}}^p$ proportional to the deviatoric stress $\bm{\sigma}'$, a self contained constitutive model cannot be obtained due to the term $\dot{\eta}^p$. A viscoplastic formulation can be used to relate the equivalent plastic strain rate $\dot{\varepsilon}^p$ to the effective stress $\sigma_e$ (rather than its rate), such that,
\begin{equation}\label{eq:EpVisco}
\dot{\varepsilon}^{p} = \dot{\varepsilon} \left [\frac{\sigma_e}{\sigma_{ref} \sqrt{f^{2}(\varepsilon^{p})+\ell_p \eta^{p}}} \right]^{m} \, .
\end{equation}

\noindent Note that (\ref{eq:EpVisco}) differs from a standard viscoplastic model in that the reference strain rate $\dot{\varepsilon}_0$ has been replaced by the effective strain rate $\dot{\varepsilon}=\sqrt{(2/3)\bm{\varepsilon}' : \bm{\varepsilon'}}$, which facilitates attaining the rate-independent limit \cite{Kok2002}. As shown by Huang \textit{et al.} \cite{Huang2004a}, the rate-independent result is well approximated for values of $m$ equal or larger than 5. A magnitude of $m=5$ is adopted throughout this work.\\

The first-order version of MSG plasticity, so-called conventional mechanism-based strain gradient (CMSG) plasticity model, incorporates gradient effects through the incremental plastic modulus. The balance equation (\ref{eq:balance}a) is identical to that of the conventional plasticity theory and changes are implemented in the computation of the material Jacobian $\bm{C}_{ep}$ and, consequently, of the stress tensor $\bm{\sigma}$. The equations relating the rate of the stress tensor with the volumetric strain rate $\text{tr} (\dot{\bm{\varepsilon}})$ and the deviatoric strain rate $\dot{\bm{\varepsilon}}'$ are identical to conventional plasticity. Thus, for a bulk modulus $K$ and a shear modulus $\mu$,
\begin{equation}
    \text{tr} (\dot{\bm{\varepsilon}}) = \frac{\text{tr} (\dot{\bm{\sigma}})}{3K} \, ,
\end{equation}
\begin{equation}\label{eq:CMSGdev}
    \dot{\bm{\varepsilon}}' = \frac{\dot{\bm{\sigma}}}{2 \mu} + \frac{3 \dot{\varepsilon}^p}{2 \sigma_e} \, .
\end{equation}

Re-arranging Eqs. (\ref{eq:EpVisco})-(\ref{eq:CMSGdev}) one can obtain the material Jacobian. Thus, the rate of the stress tensor reads
\begin{equation}
\dot{\bm{\sigma}}=K \textnormal{tr} \left( \dot{ \bm{\varepsilon}} \right) \bm{I}+2\mu \left\{ \dot{\bm{\varepsilon}}' - \frac{3\dot{\varepsilon}}{2\sigma_e}\left[\frac{\sigma_e}{\sigma_{ref} \sqrt{f^{2}(\varepsilon^{p})+\ell_p \eta^{p}}} \right]^{m} \dot{\bm{\sigma}}' \right\} \, .
\end{equation}

To compute the effective strain gradient $\eta^p$, we first interpolate  the components of the plastic strain tensor $\bm{\varepsilon}^p$ within the element and then differentiate the shape functions. 

\subsection{Addressing irreversibility and crack growth in compression}
\label{Sec:IrreversibilitySplit}

Damage must be an irreversible process,
\begin{equation}
\dot{\phi} \geq 0 \, ,
\end{equation}

\noindent and constraints should be defined to enforce this. Here, we choose to define a history field variable $\mathcal{H}$ \cite{Miehe2010a} to ensure damage irreversibility. Since the effective plastic work is assumed to increase monotonically, the history field variable only relates to the elastic contribution to fracture, such that the following Karush-Kuhn-Tucker (KKT) conditions are satisfied:
\begin{equation}
     \psi^e - \mathcal{H} \leq 0 \text{,} \hspace{8mm} \dot{\mathcal{H}} \geq 0 \text{,} \hspace{8mm} \dot{\mathcal{H}}(\psi_e-\mathcal{H})=0 \, \, .
    \centering
\end{equation}

\noindent Accordingly, for a current time $t$, over a total time $t_t$, the history field can be defined as,
\begin{equation}\label{eq:H}
    \mathcal{H} = \text{max}_{t \in [0, t_t]} \, \psi^e \left( t \right) \, .
\end{equation}

Moreover, we introduce a decomposition of the elastic strain energy density $\psi^e$ to prevent cracking in compressive strain states. Specifically, we use the volumetric-deviatoric split proposed by Amor \textit{et al.} \cite{Amor2009}, such that $\psi^e$ is decomposed into tensile $\psi^e_+$ and compressive $\psi^e_-$ terms, which are defined as follows:
\begin{equation}
    \psi^e_+ = \frac{1}{2} \lambda \langle \text{tr} \left( \bm{\varepsilon} \right) \rangle_+^2 + \mu \left( \bm{\varepsilon}' : \bm{\varepsilon}' \right)
\end{equation}
\begin{equation}
    \psi^e_-= \frac{1}{2} \lambda \langle \text{tr} \left( \bm{\varepsilon} \right) \rangle_-^2
\end{equation}

\noindent where $\langle \, \rangle$ denotes the Macaulay brackets. The present volumetric-deviatoric split is implemented using a so-called \emph{hybrid} approach \cite{Ambati2015}. That is, the split is not considered in the balance of linear momentum and is only taken into consideration in the phase field evolution law. Accordingly, the history field is based on the largest value of $\psi^e_+$ and will be referred to as $\mathcal{H}^+$ henceforth. 

\subsection{Finite element discretisation}
\label{Sec:FEdiscretization}

Recall the principle of virtual work (\ref{eq:PVW}) and consider the constitutive choices outlined in Sections \ref{Sec:ConstitutiveTheory} and \ref{Sec:IrreversibilitySplit}. The weak form for the coupled deformation-phase field fracture problem reads
\begin{align}\label{Eq:weak}
  \int_{\Omega} \bigg\{ \left( 1 - \phi \right)^2  \bm{\sigma}_0 : \delta \bm{\varepsilon} &  -2(1-\phi)\delta \phi \, \left(\mathcal{H}^+ + \psi^p \right) \nonumber \\ 
       &  +G_c \left( \frac{\phi}{\ell_f} \delta \phi
       + \ell_f \nabla \phi \cdot \nabla \delta \phi \right) \bigg\}  \, \mathrm{d}V = 0 \, ,
\end{align}

\noindent where $\bm{\sigma}_0$ is the undamaged stress tensor. Now, adopting Voigt notation, consider the following finite element interpolation for the nodal variables: the displacement vector $\bm{u}$ and the phase field $\phi$,
\begin{equation}\label{Eq:Discretization}
    \bm{u}=\sum_{i=1}^n \bm{N}_i \bm{u}_i \, , \,\,\,\,\,\,\,\,\,  \phi=\sum_{i=1}^n N_i \phi_i
\end{equation}

\noindent Here, $n$ is the number of nodes and $\bm{N}_i$ are the interpolation matrices - diagonal matrices with the nodal shape functions $N_i$ as components. Similarly, the corresponding gradient quantities are discretised as follows,
\begin{equation}
    \bm{\varepsilon}=\sum_{i=1}^n \bm{B}^u_i \bm{u}_i \, , \,\,\,\,\,\,\,\,\,  \nabla \phi=\sum_{i=1}^n \bm{B}_i \phi_i\, ,
\end{equation}

\noindent where $\bm{B}_i$ are vectors with the spatial derivatives of the shape functions and $\bm{B}^u_i$ denotes the standard strain-displacement matrices. Now, making use of this discretisation, and considering that (\ref{Eq:weak}) must hold for arbitrary values of the primal kinematic variables, the residuals can be derived as follows: 
\begin{equation} \label{eq:residualStagU}
    \bm{r}_{i}^u =\int_\Omega \left\{ \left[(1-\phi)^{2}+ \kappa \right] {(\bm{B}_{i}^u)}^{T} \bm{\sigma}_0 \right\} \, \mathrm{d}V
\end{equation}
\begin{equation}
    r_{i}^{\phi}= \int_\Omega \left[ -2(1-\phi) N_{i}  \left( \mathcal{H}^+ + \psi^p  \right) + G_c \left(\dfrac{\phi}{\ell_f} N_{i}  + \ell_f \bm{B}_{i}^T \nabla \phi \right) \right] \, \mathrm{d}V
\end{equation}

\noindent with $\kappa$ being a sufficiently small numerical parameter introduced to keep the system of equations well-conditioned when $\phi=1$. We choose to adopt a value of $\kappa=1 \times 10^{-7}$ in this work. The components of the stiffness matrices can then be obtained by differentiating the residuals with respect to the incremental nodal variables as follows:
\begin{equation}\label{Eq:Ku}
    \bm{K}_{ij}^{\bm{u}} = \frac{\partial \bm{r}_{i}^{\bm{u}}}{\partial \bm{u}_{j}} = \int_\Omega \left\{ \left[(1-\phi)^2+ \kappa \right] {(\bm{B}_i^{\bm{u}})}^T \bm{C}_{ep} \bm{B}_j^{\bm{u}} \right\} \, \mathrm{d}V  \, ,
\end{equation}
\begin{equation}
    \bm{K}_{ij}^\phi = \dfrac{\partial r_{i}^{\phi}}{\partial \phi_{j}} = \int_\Omega \left\{ \left[ 2 \left( \mathcal{H}^+ + \psi^p \right) + \frac{G_c}{\ell_f} \right]  N_{i} N_{j} + G_c \ell_f \bm{B}_i^T \bm{B}_j  \right\} \, \mathrm{d}V  \, .
\end{equation}

The linearised finite element system is solved in an incremental manner, using the Newton-Raphson method. The solution scheme follows a so-called \emph{staggered} approach \cite{Miehe2010a}, in that the solutions for the displacement and phase field problems are obtained sequentially. While working on this manuscript, two works have appeared showing that quasi-Newton solution schemes can be used to enable robust and efficient (unconditionally stable) monolithic implementations \cite{Wu2020a,TAFM2020}. The use of quasi-Newton schemes will be the aim of future endeavours. 

\section{Results}
\label{Sec:Results}

We proceed to showcase the capabilities of the model in predicting elastic-plastic fracture and investigate the interplay between the plastic and fracture length scales. Firstly, we use a boundary layer model to conduct a parametric study and estimate the influence of different conditions on the crack growth resistance (Section \ref{Sec:Rcurves}). Secondly, in Section \ref{eq:CT}, crack propagation is modelled in a Compact Tension experiment. Finally, crack nucleation and subsequent failure is predicted in an asymmetric double-notched specimen (Section \ref{sec:asymmetricbar}).

\subsection{Crack growth resistance curves (R-curves)}
\label{Sec:Rcurves}

We investigate the role of fracture and plastic length scale parameters on the fracture resistance by prescribing a remote mode I elastic $K_I$-field. Under small scale yielding conditions, the stress state in a cracked solid is characterised by the stress intensity factor $K_I$. Thus, the crack growth resistance can be characterised by predicting the crack extension $\Delta a$ as a function of the remote $K_I$, in what is usually referred to as crack growth resistance curves or R-curves. The remote $K_I$-field can be prescribed using the William's solution \cite{Williams1957} and prescribing the displacement of the outer nodes of the model. Considering both a polar $(r, \theta)$ and a Cartesian $(x,y)$ coordinate system centred at the crack tip, with the crack plane along the negative $x$-axis, the displacement field associated with a given value of $K_I$ reads,
\begin{equation}\label{eq:U_Ibl}
    u_i = \frac{K_I}{E} r^{1/2} f_i \left( \theta, \nu \right)
\end{equation}

\noindent where $E$ is Young's modulus, $\nu$ denotes Poisson's ratio, the subscript index $i$ equals $x$ or $y$, and the functions $f_i \left( \theta , \nu \right)$ are given by
\begin{equation}
    f_x = \frac{1+\nu}{\sqrt{2 \pi}} \left( 3 - 4 \nu - \cos \theta \right) \cos \left( \frac{\theta}{2} \right)
\end{equation}
\begin{equation}\label{eq:FU_Ibl}
    f_y = \frac{1+\nu}{\sqrt{2 \pi}} \left( 3 - 4 \nu - \cos \theta \right) \sin \left( \frac{\theta}{2} \right)
\end{equation}

\begin{figure}[H]
  \makebox[\textwidth][c]{\includegraphics[width=0.95\textwidth]{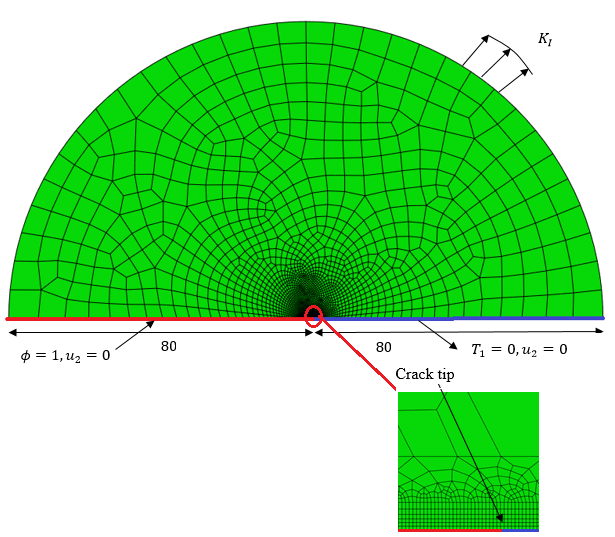}}%
  \caption{Crack growth resistance: geometry (in mm), loading configuration for the boundary layer formulation and detail of the finite element mesh.}
  \label{fig:SketchBoundaryLayer}
\end{figure}

We consider a circular solid and prescribe the magnitude of the displacement field in the outer boundary in agreement with (\ref{eq:U_Ibl})-(\ref{eq:FU_Ibl}). As shown in Fig. \ref{fig:SketchBoundaryLayer}, only the upper half of the model is considered due to symmetry. The crack is introduced by defining the initial value of the phase field equal to one along the crack plane: $\phi (t=0)=1$. The finite element mesh is refined in the crack extension region, with the characteristic element length being in all cases 5 times smaller than the phase field length scale, to ensure mesh insensitive results \cite{CMAME2018,Mandal2019}. The model is discretised with a total of 10,550 quadratic quadrilateral elements with reduced integration and three degrees-of-freedom per node ($u_x$, $u_y$, $\phi$). 

Fracture is simulated in a solid with the following material properties: $\sigma_Y/E=0.003$, Poisson's ratio $\nu=0.3$, and strain hardening exponent $N=0.2$. Inspired in the cohesive zone modelling work by Tvergaard and Hutchinson \cite{Tvergaard1992}, we define a reference stress intensity factor as,
\begin{equation}
K_0 = \left( \frac{E G_c}{1-\nu^2} \right)^{1/2}    
\end{equation}

\noindent and fracture process zone length as,
\begin{equation}\label{eq:R0}
    R_0 = \frac{1}{3 \pi \left( 1 - \nu^2 \right)} \frac{E G_c}{\sigma_Y^2} \, .
\end{equation}

A relation between the non-dimensional group $\ell_f/R_0$ and the material strength can be established by considering both (\ref{eq:effectiveS}) and (\ref{eq:R0}), such that
\begin{equation}
    \frac{R_0}{\ell_f} = \frac{256}{81 \pi \left( 1 - \nu^2 \right)} \left( \frac{\hat{\sigma}}{\sigma_Y} \right)^2 \approx \left( \frac{\hat{\sigma}}{\sigma_Y}\right)^2\, .
\end{equation}

\begin{figure}[H]
  \makebox[\textwidth][c]{\includegraphics[width=1.1\textwidth]{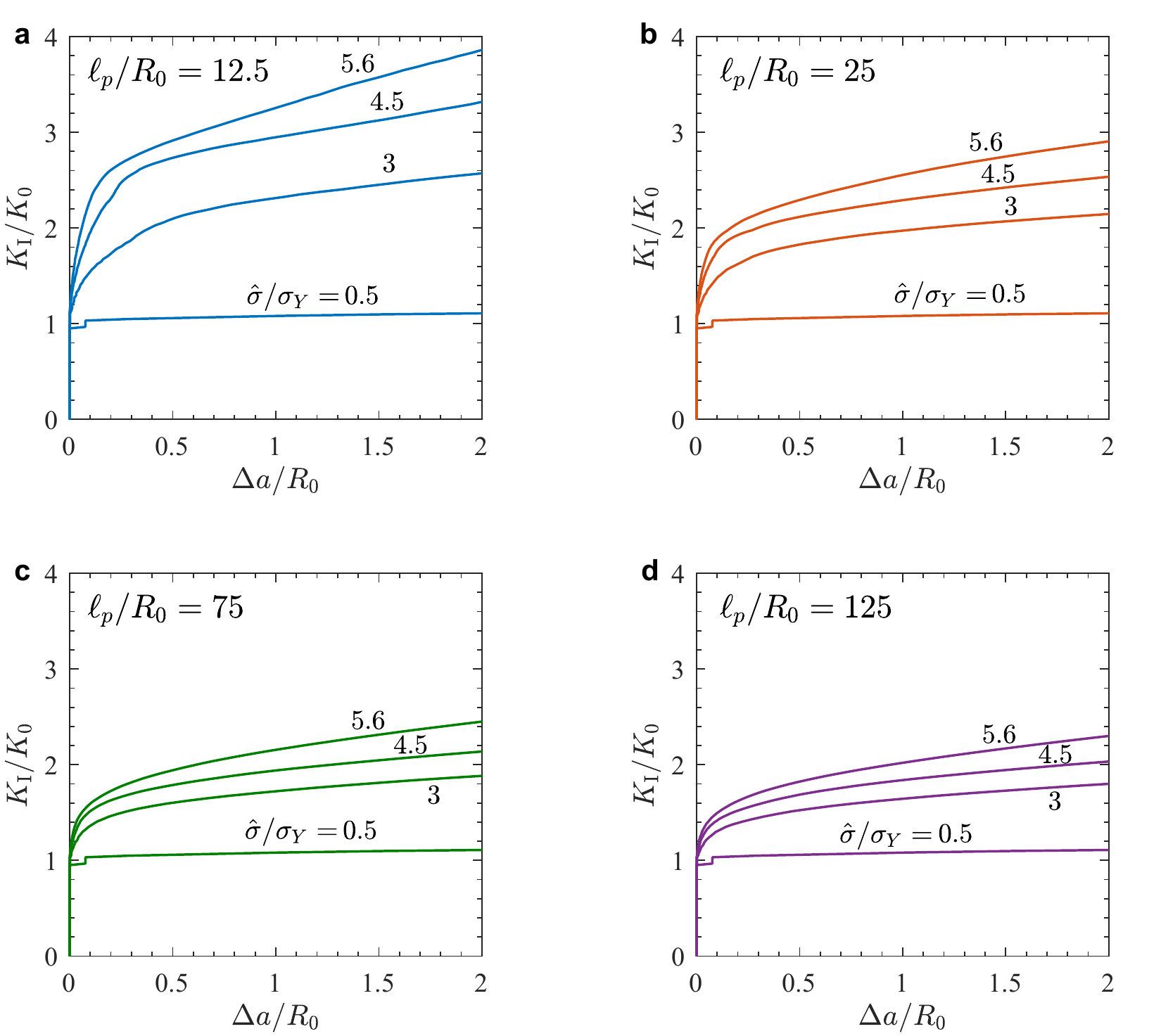}}%
  \caption{Crack growth resistance: influence of the plastic length scale $\ell_p$: (a) $\ell_p/R_0=12.5$, (b) $\ell_p/R_0=25$, (c) $\ell_p/R_0=75$, and (d) $\ell_p/R_0=125$.}
  \label{fig:RcurveLp}
\end{figure}

\begin{figure}[H]
  \makebox[\textwidth][c]{\includegraphics[width=1.1\textwidth]{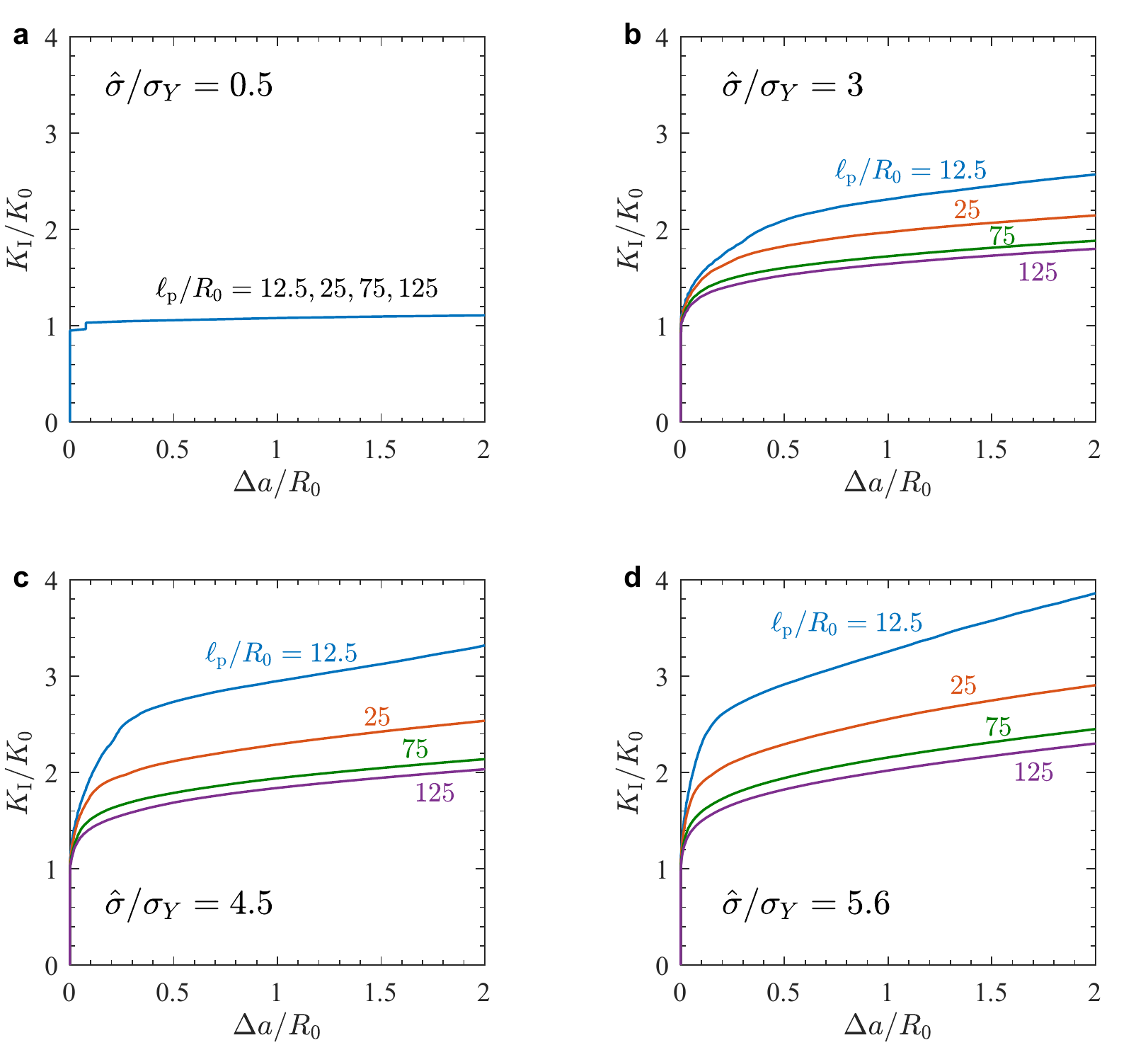}}%
  \caption{Crack growth resistance: influence of the material strength $\hat{\sigma}$ (phase field length scale $\ell_f$): (a) $\hat{\sigma}/\sigma_Y=0.5$, (b) $\hat{\sigma}/\sigma_Y=3$, (c) $\hat{\sigma}/\sigma_Y=4.5$, and (d) $\hat{\sigma}/\sigma_Y=5.6$.}
  \label{fig:RcurveLf}
\end{figure}

\noindent Thus, the non-dimensional group $R_0/\ell_f$ governs the material strength, which will influence the dissipation taking place with crack growth. Another relevant non-dimensional set is $\ell_p/R_0$, governing the capacity of the material to exhibit additional hardening due to the presence of plastic strain gradients and GNDs. We shall investigate the role that both $\ell_f/R_0$ and $\ell_p/R_0$ have on the material response. The results obtained are shown in Figs. \ref{fig:RcurveLp} and \ref{fig:RcurveLf}, in terms of the normalised applied $K_I/K_0$ versus the normalised crack extension $\Delta a /R_0$, for different $\ell_f/R_0$ and $\ell_p/R_0$ ratios. 

Each of the sub-figures of Fig. \ref{fig:RcurveLp} shows crack growth resistance curves for four selected $\hat{\sigma}/\sigma_Y$ ratios for a given plastic length scale value: (a) $\ell_p/R_0=12.5$, (b) $\ell_p/R_0=25$, (c) $\ell_p/R_0=75$, and (d) $\ell_p/R_0=125$. For all values of $\ell_p/R_0$, a flat R-curve is predicted when $\hat{\sigma}/\sigma_Y=0.5$. This is in agreement with expectations, as there is no plastic dissipation if the yielding stress is not reached. Larger values of $\hat{\sigma}/\sigma_Y$ exhibit a rising R-curve due to inelastic dissipation. We also observe that the initiation of crack growth occurs at $K_I \approx K_0$ for all $\hat{\sigma}/\sigma_Y$ and $\ell_p/R_0$ combinations. Again, this is in agreement with expectations - in the presence of a large crack (toughness-driven failure), phase field models are able to capture crack initiation at the appropriate energy release rate \cite{PTRSA2021}. Comparing results across sub-figures, it can be readily seen that larger values of $\ell_p/R_0$ reduce the degree of toughening associated with plastic dissipation. This is more clearly observed in Fig. \ref{fig:RcurveLf}, where the different $\ell_p/R_0$ curves are shown for specific choices of $\hat{\sigma}/\sigma_Y$. A larger $\ell_p/R_0$ ratio implies a greater influence of plastic strain gradients and GNDs, which translates into higher crack tip stresses that facilitate crack growth. It is worth emphasising that cracking is observed for $\hat{\sigma}/\sigma_Y$ values that are significantly larger than the strengths at which failure is precluded when using the conventional plasticity theory. In the absence of gradient effects, crack tip stresses are not high enough to trigger crack growth for strength values equal or larger than $\hat{\sigma}=4.5 \sigma_Y$ for a material with $N=0.2$ and $\sigma_Y/E=0.003$ \cite{Tvergaard1992,Wei1997,JMPS2019}. The low crack tip stresses attained with conventional plasticity theories are at odds with the observations of brittle fracture in the presence of plasticity, as in low temperature cleavage of ferritic steels \cite{EJMAS2019b}, the failure of bi-material interfaces or hydrogen embrittlement \cite{AM2016}. Brittle interfaces have strengths on the order of $\hat{\sigma}=10 \sigma_Y$, which can only be reached if the local strengthening effect of crack tip plastic strain gradients is accounted for.

\subsection{Compact Tension sample}
\label{eq:CT}

\begin{figure}[H]
  \makebox[\textwidth][c]{\includegraphics[width=1\textwidth]{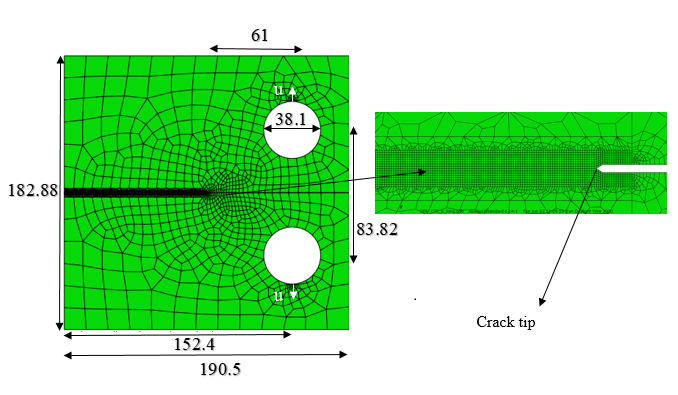}}%
  \caption{Compact Tension sample: geometry (in mm), loading configuration and finite element mesh.}
  \label{fig:SketchCT}
\end{figure}

We shall now simulate a Compact Tension fracture experiment. The geometry is shown in Fig. \ref{fig:SketchCT}, with the dimensions given in mm. The load is applied by prescribing the vertical displacement of the nodes in the pin holes, which are not allowed to move in the horizontal direction. We assume a material with Young's modulus $E=71.48$ GPa, Poisson's ratio $\nu=0.3$, initial yield stress $\sigma_Y=345$ MPa and strain hardening exponent $N=0.2$. The material toughness is chosen to be equal to $G_c=9.31$ MPa$\cdot$mm and the phase field length scale equals $\ell_f=0.15$ mm. Accordingly, the characteristic element size along the crack propagation region is chosen to be of 0.03 mm. The finite element mesh employed is shown in Fig. \ref{fig:SketchCT}; approximately 28,000 quadratic quadrilateral elements have been employed. 

\begin{figure}[H]
  \makebox[\textwidth][c]{\includegraphics[width=1\textwidth]{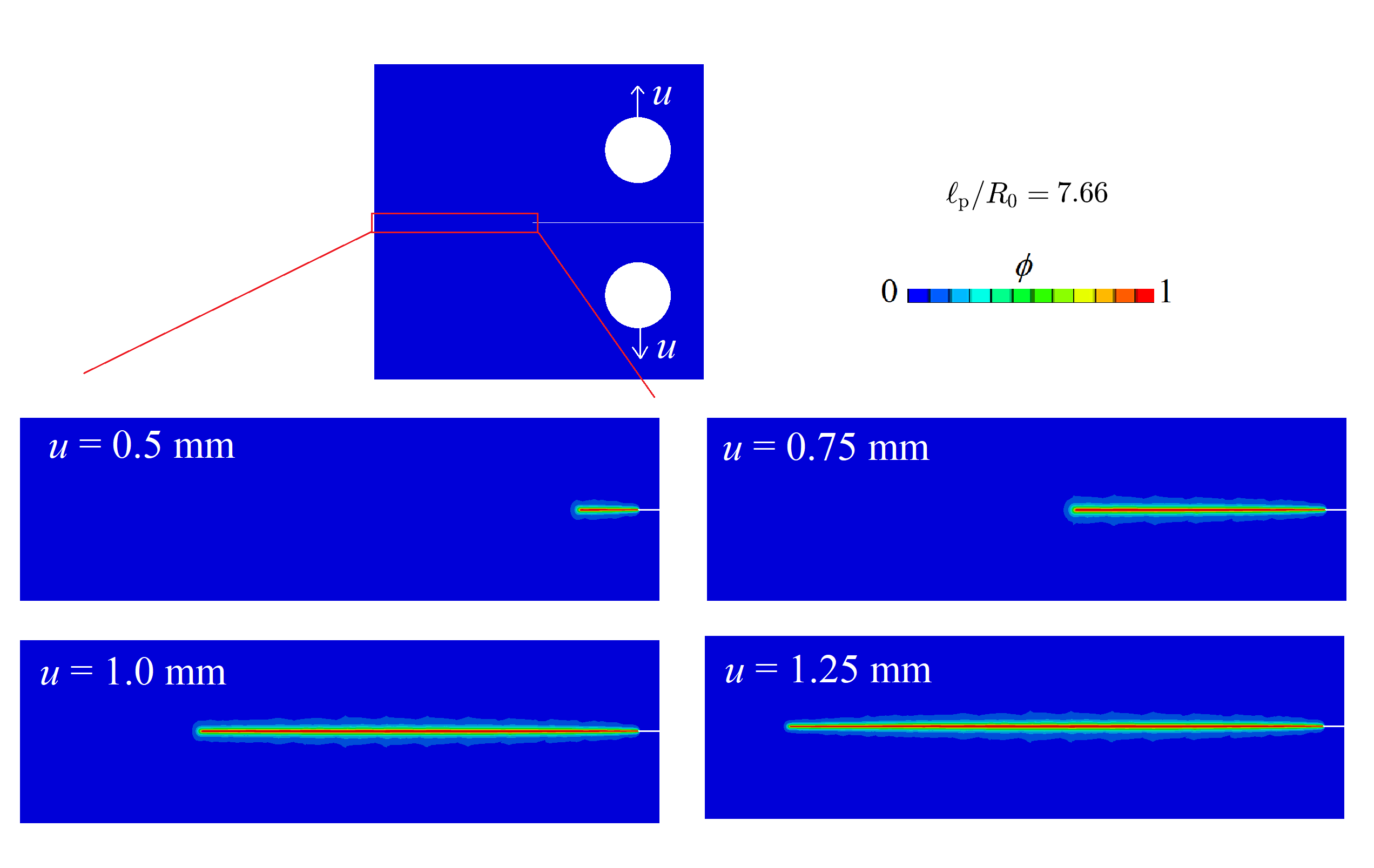}}%
  \caption{Compact Tension sample: phase field $\phi$ contours for selected load levels, as characterised by the applied displacement $u$.}
  \label{fig:CT_Contours}
\end{figure}

The phase field contours for different levels of the applied displacement are shown in Fig. \ref{fig:CT_Contours}. The same qualitative trend is observed for both the conventional plasticity case ($\ell_p/R_0=0$) and the gradient-enhanced analysis ($\ell_p/R_0>0$). A mode I crack starts growing from the tip of the initial defect and propagates all the way up to the final failure of the Compact Tension specimen. Given the differences between the applied displacement $u$ values reported in each sub-figure, it can be observed that crack growth takes place in a stable manner.

\begin{figure}[H]
  \makebox[\textwidth][c]{\includegraphics[width=0.7\textwidth]{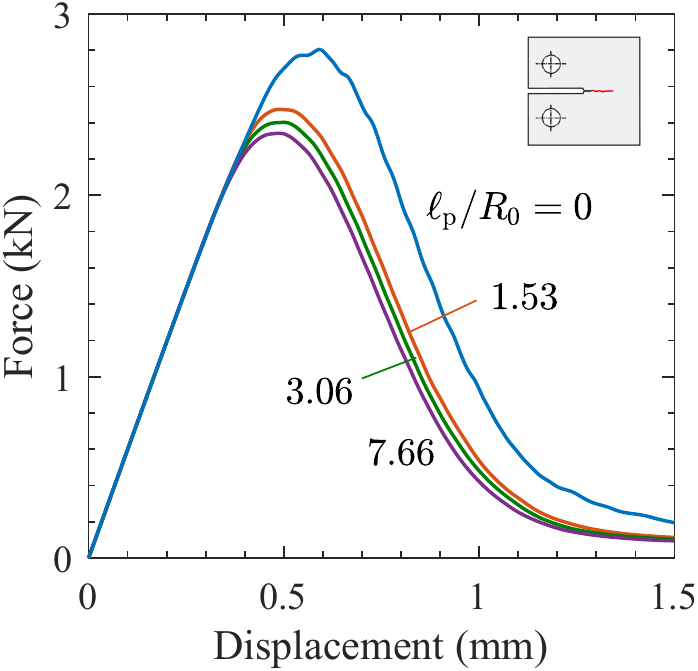}}%
  \caption{Compact Tension experiment: force versus displacement response obtained for selected values of the plastic length scale $\ell_p/R_0$.}
  \label{fig:CT_ForceDisp}
\end{figure}

The force versus displacement curves obtained for selected values of the plastic length scale $\ell_p$ are shown in Fig. \ref{fig:CT_ForceDisp}. In all cases, the force increases with the applied displacement, up to a maximum value located within the 2.3-2.8 kN range, and then drops in a rather smooth manner, as a result of the stable crack propagation observed. Increasing the magnitude of the plastic length scale translates into higher crack tip stresses, which facilitate fracture: the peak load decreases with increasing $\ell_p/R_0$. The softening part of the force versus displacement response exhibits a similar qualitative trend for all $\ell_p/R_0$ values.

\begin{figure}[H]
  \makebox[\textwidth][c]{\includegraphics[width=0.8\textwidth]{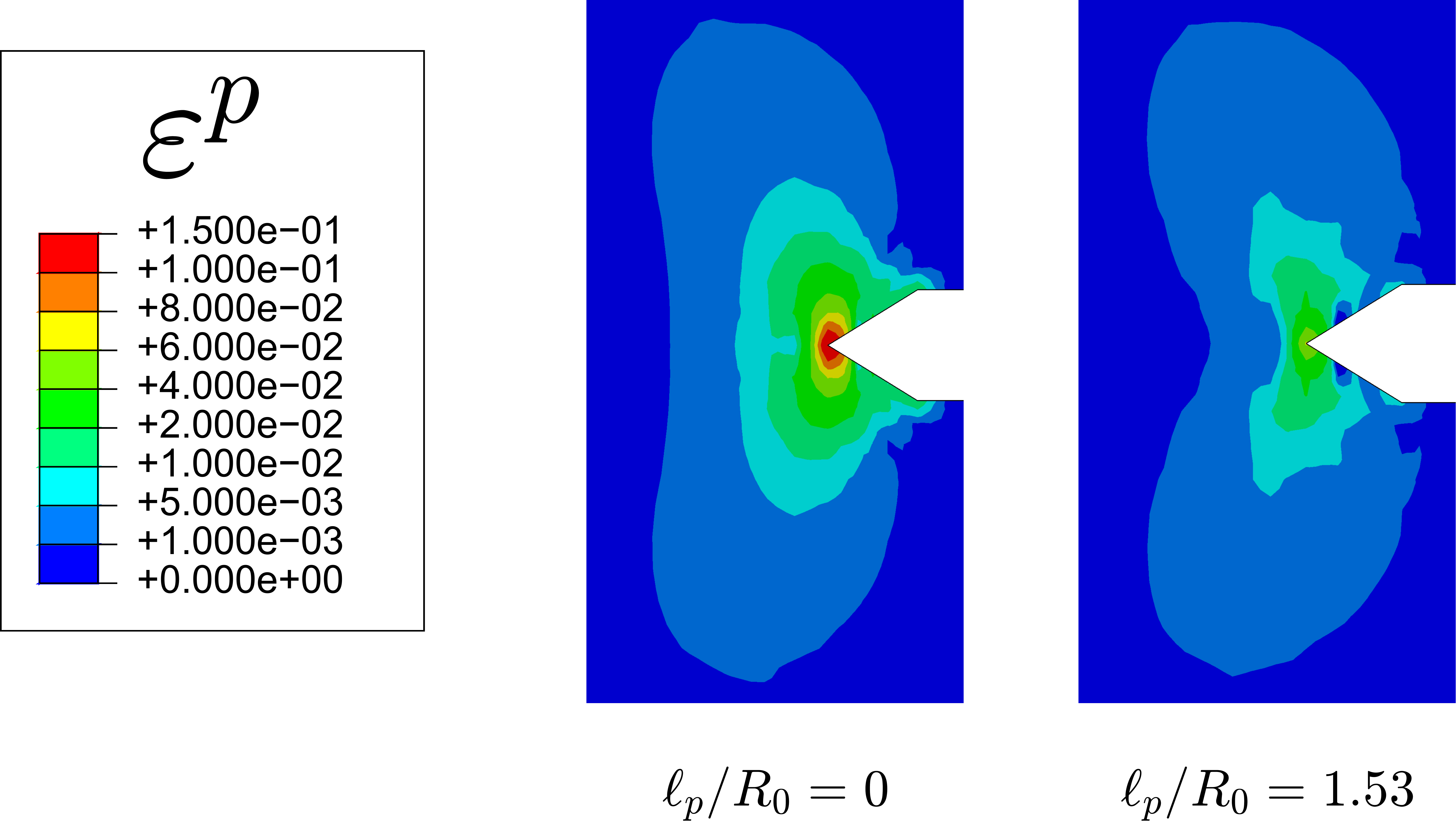}}%
  \caption{Compact Tension experiment: equivalent plastic strain $\varepsilon^p$ shortly before the onset of crack growth. Results are shown for both conventional ($\ell_p/R_0=0$) and gradient-enhanced plasticity ($\ell_p/R_0=1.53$).}
  \label{fig:CT_ContoursPEEQ}
\end{figure}

Finally, Fig. \ref{fig:CT_ContoursPEEQ} shows the contours of equivalent plastic strain for the cases of $\ell_p/R_0=0$ and $\ell_p/R_0=1.53$, shortly before the onset of crack growth. It can be observed that the shape and size of the plastic zone is similar for both conventional and gradient-enhanced plasticity. However, the local hardening resulting from large plastic strain gradients at the crack tip significantly reduces the crack tip plasticity levels in the $\ell_p/R_0>0$ case. The maximum $\varepsilon^p$ values observed in the case of conventional plasticity exceed 10\%, indicating that a finite strain analysis would provide a more precise description in such a case.

\subsection{Asymmetric double-notch specimen}
\label{sec:asymmetricbar}

\begin{figure}[H]
  \makebox[\textwidth][c]{\includegraphics[width=0.6\textwidth]{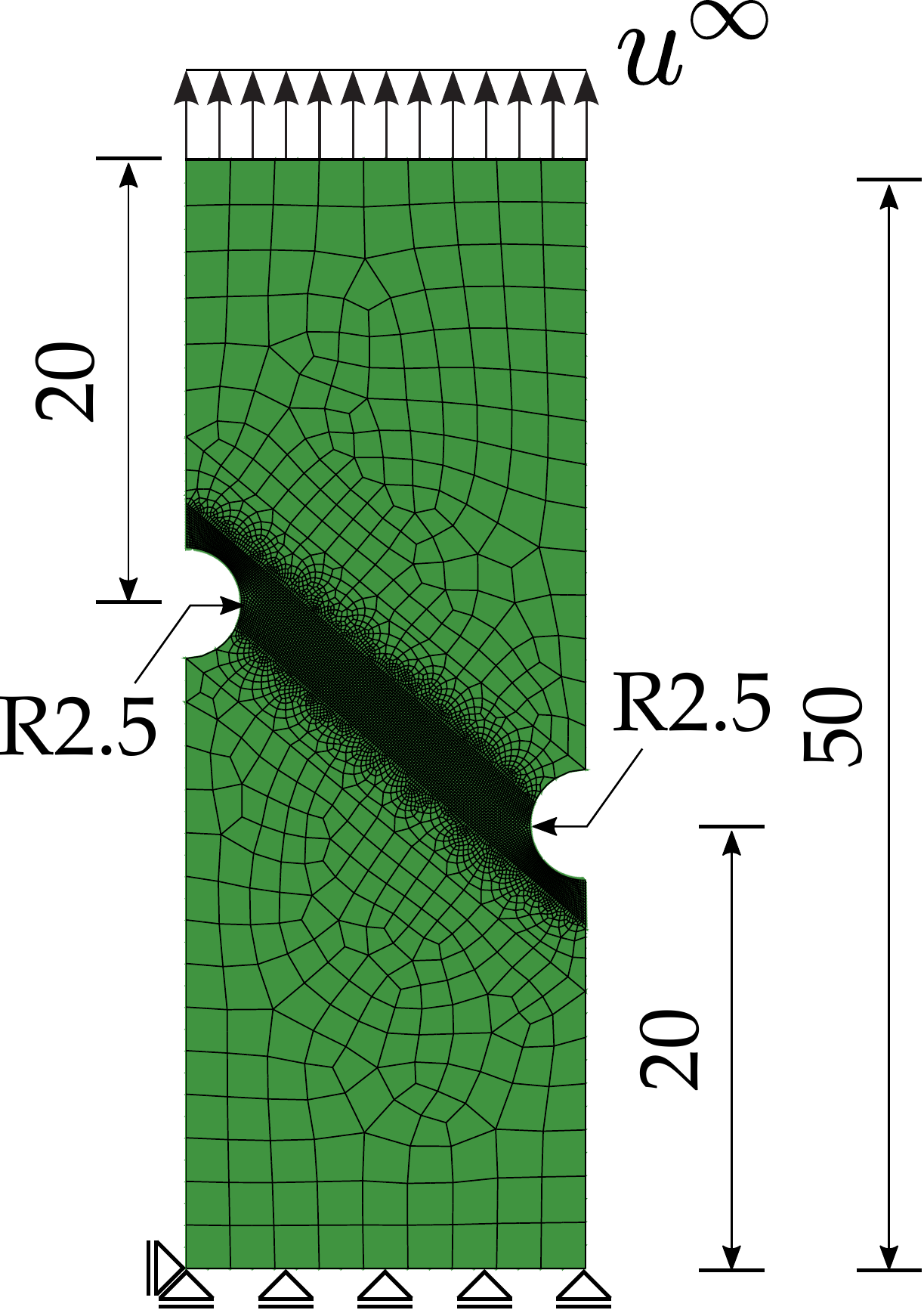}}%
  \caption{Asymmetric double-notch specimen: geometry (in mm), loading configuration and finite element mesh.}
  \label{fig:SketchAsy}
\end{figure}

We investigate the capabilities of the model for predicting mixed-mode fracture and the coalescence of cracks by simulating fracture in an asymmetrically notched plane strain bar. The geometry (in mm), finite element mesh and loading configurations are shown in Fig. \ref{fig:SketchAsy}. Two notches of 2.5 mm radii are present at each side of the sample. The bottom edge of the sample has its vertical displacement constrained while a remote vertical displacement $u$ is applied at the top edge. To prevent rigid body motion, the bottom-left corner has its horizontal displacement constrained. Following the conventional plasticity study by Fang \textit{et al.} \cite{Fang2019}, the material is assumed to exhibit linear work hardening such that, instead of (\ref{eq:HardeningPowerLaw}), the conventional strain hardening behaviour is characterised by,
\begin{equation}
    \sigma = \sigma_Y \left( 1 + \frac{\varepsilon^p E_t}{\sigma_Y} \right) 
\end{equation}

\noindent where $E_t$ is the elastic-plastic tangent modulus, assumed to be equal to $E_t=714.8$ MPa. Otherwise, the material properties resemble those of the previous case study, $E=71.48$ GPa, $\nu=0.3$, and $\sigma_Y=345$ MPa. Also, the critical energy release rate and the phase field length scale respectively read $G_c=9.31$ MPa$\cdot$mm and $\ell_f=0.15$ mm. As shown in Fig. \ref{fig:SketchAsy}, the mesh is refined in the potential crack propagation region to ensure that $\ell_f$ is resolved; the characteristic element length is at least 5 times smaller than $\ell_f$. 

\begin{figure}[H]
  \makebox[\textwidth][c]{\includegraphics[width=0.7\textwidth]{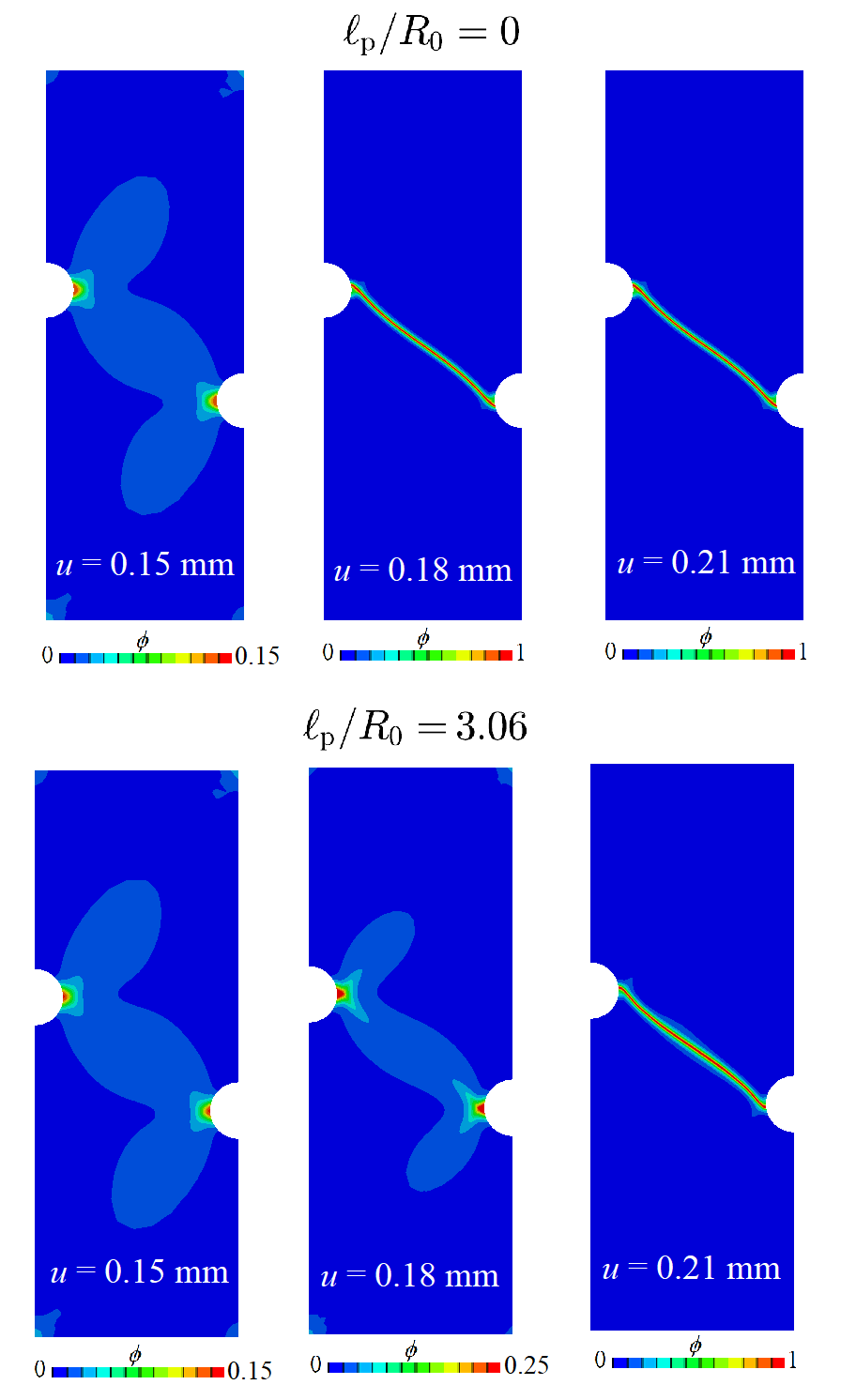}}%
  \caption{Asymmetric double-notch specimen: phase field $\phi$ contours for selected load levels, as characterised by the applied displacement $u$. Results are shown for both conventional plasticity ($\ell_p=0$) and the present dislocation-based gradient plasticity model with $\ell_p/R_0=3.06$.}
  \label{fig:Asy_Contours}
\end{figure}

The damage contours, as characterised by the phase field order parameter, are shown in Fig. \ref{fig:Asy_Contours} for different values of the applied displacement. Results are shown for both the conventional plasticity case ($\ell_p=0$) and for the mechanism-based formulation presented in Section \ref{Sec:Theory}, with $\ell_p/R_0=3.06$. In both cases, it can be observed from the $u$ values that failure occurs in a rather sudden manner, with two defects nucleating from the tip of each notch and very fast coalescencing with each other. 

\begin{figure}[H]
  \makebox[\textwidth][c]{\includegraphics[width=0.7\textwidth]{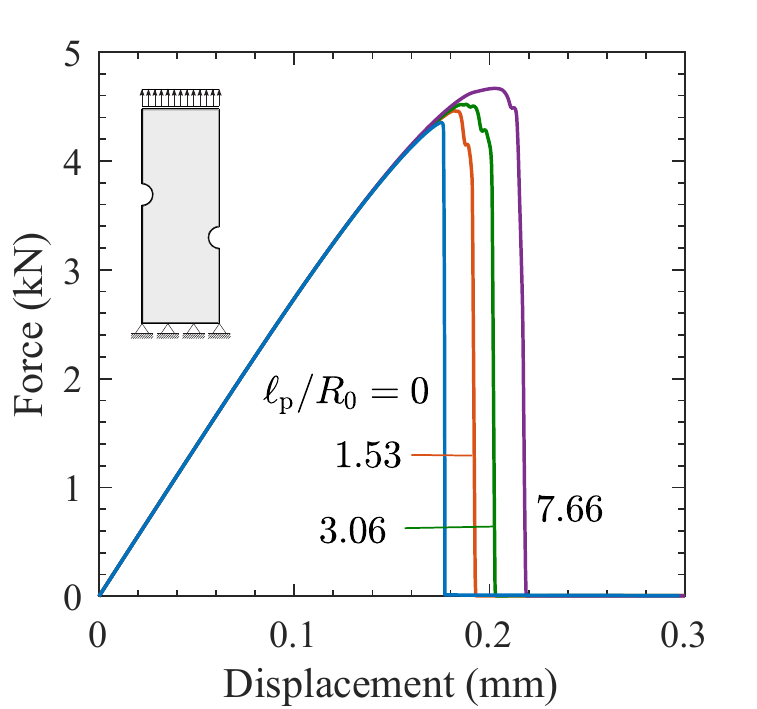}}%
  \caption{Asymmetric double-notch specimen: force versus displacement response obtained for selected values of the plastic length scale $\ell_p/R_0$.}
  \label{fig:Asy_ForceDisp}
\end{figure}

Interestingly, it appears that the case accounting for the role of plastic strain gradients would lead to a later failure. This is arguably because, in this boundary value problem, failure is driven by plastic localisation. Plastic strain gradients are less relevant ahead of blunted notches, relative to sharp defects, so it appears likely that the main role of strain gradient hardening is to delay plastic localisation. This is clearly observed in the force versus displacement response. As show in Fig. \ref{fig:Asy_ForceDisp}, increasing the magnitude of the plastic length scale further delays the localisation event, raising the maximum load. In all $\ell_p/R_0$ cases, in agreement with the damage contours, a sharp drop in the load carrying capacity is observed shortly after reaching the peak load, indicative of unstable crack growth. 

\section{Conclusions}
\label{Sec:Concluding remarks}

We have presented a new non-local damage formulation for elastic-plastic solids. The theoretical and computational framework presented builds upon two main pillars: (i) a dislocation-based model to capture micro-scale plastic deformation, and (ii) a phase field description of damage. The motivation behind this enriched continuum description of elastic-plastic deformation is to incorporate the role that plastic strain gradients (and geometrically necessary dislocations) play in promoting strain hardening and elevating the stresses ahead of cracks and other sharp defects. The phase field model provides a suitable energy-based framework for simulating the onset and evolution of damage without the limitations of discrete approaches. The non-locality of the plasticity and damage formulations provides a regularised framework for softening and damage, and results in the existence of two length scales in the constitutive theory: a plastic length scale $\ell_p$ that can be calibrated with micro-scale experiments, and a fracture length scale $\ell_f$, which governs the strength of the solid.\\

The non-local plastic-damage formulation presented is numerically implemented using the finite element method. Several numerical experiments are conducted to investigate the predictions of the model in a wide range of scenarios: small and large scale yielding, sharp and blunted defects, linear and power-law hardening behaviour. We find that, in agreement with expectations, decreasing the phase field length scale results in a strength elevation, which translates into a large degree of plastic dissipation during the crack propagation process; i.e., a higher crack growth resistance. The role of the plastic length scale is more difficult to anticipate. In the presence of a sharp crack, large gradients in plastic strain exist locally, elevating crack tip stresses much beyond the predictions of conventional plasticity. This results in a significant reduction of crack growth resistance and a smaller steady-state toughness. Failure is observed at high material strengths, rationalising brittle fracture in the presence of plasticity. However, in scenarios where failure is driven by the localisation of plastic flow, plastic strain gradients appear to delay fracture. The additional hardening resulting from the extra storage of dislocations reduces the plastic work, for the same load level, relative to conventional von Mises plasticity. This behaviour is likely to be related to the choice of a fracture driving force based on the \emph{total} strain energy density (elastic and plastic). Notwithstanding, it appears sensible to consider, to a certain extent, the plastic work in the damage process when simulating ductile fracture mechanisms.  
 
\section{Acknowledgements}
\label{Acknowledge of funding}

 E. Mart\'{\i}nez-Pa\~neda acknowledges financial support from the EPSRC (grant EP/V009680/1) and from the Royal Commission for the 1851 Exhibition (RF496/2018).




\bibliographystyle{elsarticle-num}
\bibliography{library}



\end{document}